\documentstyle[12pt,aaspp4,psfig,natbib209]{article}

\def\la{\mathrel{\hbox{\rlap{\hbox{\lower4pt\hbox{$\sim$}}}\hbox{$<$}}}}
\def\ga{\mathrel{\hbox{\rlap{\hbox{\lower4pt\hbox{$\sim$}}}\hbox{$>$}}}}

\newcommand{\be}{\begin{eqnarray}}
\newcommand{\ee}{\end{eqnarray}}

\newcommand{\msol}{\ifmmode{{\rm M}_\odot}\else{M$_\odot$}\fi}
\newcommand{\foe}{\ifmmode{10^{51}}\else{$10^{51}$}\fi}

\newcommand{\xni}{\ifmmode{{\rm X}_{\rm Ni}}\else{X$_{\rm Ni}$}\fi}
\def\ang{\hbox{\AA}}
\def\Teff{\ifmmode{T_{\rm eff}}\else{\hbox{$T_{\rm eff}$} }\fi}
\def\Rzero{\ifmmode{R_0}\else{\hbox{$R_0$} }\fi}

\def\SP2{{\tt IBM SP2}}

\def\PC2{{\tt PC$^2$}}

\def\logg{\log(g)}
\def\mh{[{\rm M/H}]}

\def\inu{\ifmmode{I_{\nu}}\else{\hbox{$I_{\nu}$} }\fi}
\def\snu{\ifmmode{S_{\nu}}\else{\hbox{$S_{\nu}$} }\fi}
\def\jnu{\ifmmode{J_{\nu}}\else{\hbox{$J_{\nu}$} }\fi}

\def\etal{et al.}
\def\fep{\ifmmode{{\rm Fe II}}\else\hbox{Fe~II }\fi}

  at 12.0 true pt 

\def\etal{{et al}}

\def\atlasIX{{\tt ATLAS9}}
\def\atlasXII{{\tt ATLAS12}}
\def\phoenix{{\tt PHOENIX}}
\def\PHOENIX{{\tt PHOENIX}}

\def\IUE{{\bf IUE}}
\def\water{{H$_2$O}}

\def\etal{{et al}}

\def\phoenix{{\tt PHOENIX}}

\def\IUE{{\bf IUE}}
\def\water{{H$_2$O}}

\def\mtwater{MT-\water}
\def\hitran{{\tt HITRAN92}}

\def\b{\beta}

\def\rout{\ifmmode{r_{\rm out}}\else\hbox{$r_{\rm out}$}\fi}
\def\tmax{\ifmmode{\tau_{\rm max}}\else\hbox{$\tau_{\rm max}$}\fi}
\def\tstd{\ifmmode{\tau_{\rm std}}\else\hbox{$\tau_{\rm std}$}\fi}
\def\vmax{\ifmmode{v_{\rm max}}\else\hbox{$v_{\rm max}$}\fi}
\def\muE{\ifmmode{\mu_{\rm E}}\else\hbox{$\mu_{\rm E}$}\fi} 
\def\pE{\ifmmode{p_{\rm E}}\else\hbox{$p_{\rm E}$}\fi} 
\def\bmax{\ifmmode{\b_{\rm max}}\else\hbox{$\b_{\rm max}$}\fi}
\def\kms{\hbox{$\,$km$\,$s$^{-1}$}}

\def\ang{\hbox{\AA}}

\def\Teff{\hbox{$\,T_{\rm eff}$} }

\def\alog#1{\times 10^{#1}}

\def\rout{\hbox{$r_{\rm out}$} }

\def\chistd{\ifmmode{\chi_{\rm std}}\else\hbox{$\chi_{\rm std}$}\fi}

\def\k{\,{\rm K}}
\def\K{\,{\rm K}}
\def\msol{$M_\odot$}
\def\foe{10^{51}}

\def\xni{{\rm X}_{\rm Ni}}

\def\lstar{\ifmmode{\Lambda^*}\else\hbox{$\Lambda^*$}\fi} 
\def\Rop{\ifmmode{[R_{ij}]}\else\hbox{$[R_{ij}]$}\fi}

\def\Rji{\ifmmode{[R_{ji}]}\else\hbox{$[R_{ji}]$}\fi}
\def\Rstar{\ifmmode{[R_{ij}^*]}\else\hbox{$[R_{ij}^*]$}\fi}

\def\Rjistar{\ifmmode{[R_{ji}^*]}\else\hbox{$[R_{ji}^*]$}\fi}
\def\DRji{\ifmmode{[\Delta R_{ji}]}\else\hbox{$[\Delta R_{ji}]$}\fi}
\def\DRij{\ifmmode{[\Delta R_{ij}]}\else\hbox{$[\Delta R_{ij}]$}\fi}

\def\ns{\ifmmode{N_{\rm s}}          
        \else\hbox{$N_{\rm s}$}\fi}

\def\mat#1{{\bf #1}}     
\def\vek#1{{#1}}         

\newcount\eqcount
\def
  \nummer{
    \global\advance\eqcount by 1
    (\the\eqcount)
  }

\def
  \numadv{
    \global\advance\eqcount by 1
  }

\def
   \numout#1{
     (\the\eqcount #1)
  }

\def\ivek#1#2{\ifmmode{\vek{I}^{#1}_{#2}}
        \else\hbox{$\vek{I}^{#1}_{#2}$}\fi}

\def\tmat#1#2{\ifmmode{\mat{t}^{#1}_{#2}}
        \else\hbox{$\mat{t}^{#1}_{#2}$}\fi}
\def\rmat#1#2{\ifmmode{\mat{r}^{#1}_{#2}}
        \else\hbox{$\mat{r}^{#1}_{#2}$}\fi}
\def\bvek#1#2{\ifmmode{\beta^{#1}_{#2}}
        \else\hbox{$\beta^{#1}_{#2}$}\fi}

\def\lp{\ifmmode{\lambda^+_\tau}           
        \else\hbox{$\lambda^+_\tau$}\fi}
\def\lm{\ifmmode\lambda^-_\tau             
        \else\hbox{$\lambda^-_\tau$}\fi}

\chardef\tilt=126
\begin{document}
\bibliographystyle{apj}

\title{The NextGen Model Atmosphere grid for $3000\le \Teff \le 10000\K$}

\author{Peter H. Hauschildt}
\affil{Dept.\ of Physics and Astronomy \& Center for Simulational Physics, 
University of Georgia, Athens, GA 30602-2451\\
Email: {\tt yeti@hal.physast.uga.edu}}
\author{France Allard}
\affil{Dept.\ of Physics, Wichita State University,
Wichita, KS 67260-0032\\
E-Mail: \tt allard@eureka.physics.twsu.edu}
\and
\author{E. Baron}
\affil{Dept. of Physics and Astronomy, University of Oklahoma,\\
440 W. Brooks, Rm 131, Norman, OK 73019-0225\\
E-Mail: \tt baron@mail.nhn.ou.edu}

\bigskip
\begin{center}
\em
Text \& Figures available at\\
\tt ftp://calvin.physast.uga.edu/pub/preprints/NextGen.ps.gz
\end{center}

\begin{abstract}

We present our NextGen Model Atmosphere grid for low mass stars for effective
temperatures larger than $3000\K$. These LTE models are calculated with the
same basic model assumptions and input physics as the VLMS part of the NextGen
grid so that the complete grid can be used, e.g., for consistent stellar
evolution calculations and for internally consistent analysis of cool star
spectra. This grid is also the starting point for a large grid of detailed NLTE
model atmospheres for dwarfs and giants (Hauschildt \etal, in preparation). The
models were calculated from $3000\K$ to $10000\K$ (in steps of $200\K$) for
$3.5 \le \logg \le 5.5$ (in steps of 0.5) and metallicities of $-4.0 \le \mh
\le 0.0$.

We discuss the results of the model calculations and compare our results to the
Kurucz 1994 grid. Some comparisons to standard stars like Vega and the Sun are
presented and compared with detailed NLTE calculations. 

\end{abstract}

\section{Introduction}

As part of our ongoing effort to calculate accurate line blanketed
stellar atmospheres we present the results our ``NextGen'' model
atmosphere grid for ``high'' effective temperatures ($3000\le\Teff\le
10000\K$). The results for 
very low mass stars (VLMS) and Brown Dwarfs involve the complication
of dust condensation and are discussed in Allard \etal, (in
preparation). All the NextGen models are computed with the same
assumptions to allow internally consistent analysis of, e.g., HR
diagrams of globular clusters or population synthesis models.

Most of the models have been calculated under the assumption of local
thermodynamic equilibrium (LTE) for the ionization/dissociation equilibria
and the level populations.  Thomson and Rayleigh scattering
are included and cause the source function to deviate from the Planck
function. The LTE assumption has been made mainly to be consistent
with the VLMS part of the NextGen grid. The LTE models are being used
as starting points for a large NLTE model grid that is currently being
constructed and that will replace the LTE model grid (Hauschildt \etal,
in preparation). The NLTE models that are currently available in the
NextGen model grid feature only a small subset of NLTE species and will
be replaced by full NLTE models.  We will discuss briefly how NLTE models
change the results for one example.

\cite{kurucz92} has calculated an extensive grid of LTE model atmospheres
with his \atlasIX\ code using Opacity Distribution Functions (ODF)
to include line blanketing. This model grid has set the standard
for a large number spectral analyses over a wide range of effective
temperatures. \cite{castelli94} have used the Kurucz~92 grid and the new
opacity sampling version of Kurucz's model atmosphere code \atlasXII\
\cite[]{kurucz93} to compute model atmospheres for Vega. They find very
good agreement over a wide range in wavelengths, improving significantly
upon earlier results. This is mostly due to Kurucz's update to his line
lists, resulting in an increase in the total line opacity resulting in
better fits to the ultraviolet spectra.

We use the Kurucz~92 grid to test our calculations at effective
temperatures larger than about $5000\K$. At lower temperatures, larger
differences are expected due to differences between the equations of
state (EOS) and the treatment of molecular opacities in \atlasXII\
and \phoenix. For higher effective temperatures, when molecules and
molecular opacities become less important, the synthetic spectra should
be in good agreement with our LTE models because we use the Kurucz atomic
line list \cite[]{cdrom1};  small differences are expected
due to different bound-free and free-free opacity sources and different
numerical methods used.

\section{Methods and Models}

For the model grid calculations, we use our multi-purpose
stellar atmosphere code \phoenix\ \cite[version
9.1,][]{fe2nova,parapap,nova_cno,parapap2}. In the calculations presented in
this paper, we will mostly use the static, plane-parallel LTE mode of
\phoenix, although we will compare LTE results with NLTE calculations
to estimate the systematic errors made by LTE analyses.  Details of the
numerical methods are given in the above references, so we do not repeat
the description here. We will, however, describe the changes in the
input physics and data compared to our earlier ``base'' and ``Extended''
model grids for very low mass stars.

Both atomic and molecular lines are treated with a direct 
opacity sampling method.  We do {\em not} use pre-computed opacity
sampling tables, but instead dynamically select the relevant LTE
background lines from master line lists at the beginning of each iteration
for every model and sum the contribution of every line within a search
window to compute the total line opacity at {\em arbitrary} wavelength
points.  The latter feature is crucial in NLTE calculations in which
the wavelength grid is both irregular and variable from iteration to
iteration due to changes in the physical conditions. This approach also
allows detailed and depth dependent line profiles to be used during
the iterations.

Although the direct line treatment seems at first glance computationally
prohibitive, it can lead to more accurate models. This is due to the fact that
the line forming regions in cool dwarfs span a huge range in pressures and
temperatures so that the line wings form in very different layers than the line
cores. Therefore, the physics of the line formation is best modeled by an
approach that treats the variation of the line profile and the level excitation
as accurately as possible. In addition, it is relatively straightforward to
write an efficient computer code for the direct opacity sampling treatment by
making use of data-locality, vectorization \& super-scalar execution, and
caches found in modern workstations and supercomputers. On many machines,
interpolation in the large tables required  by the ODF method takes more time
per wavelength point than direct opacity sampling because of the longer memory
access times compared to the execution speed of floating point instructions.
To make this method computationally more efficient, we employ modern numerical
techniques, e.g., vectorized and parallelized block algorithms with high data
locality \cite[]{parapap}, and we use high-end workstations or parallel
supercomputers for the model calculations.

In the calculations presented in this paper, we have have included a
constant statistical velocity field, $\xi = 2\kms$, which is treated like
a microturbulence.  The choice of lines is dictated by whether they are
stronger than a threshold $\Gamma\equiv \chi_l/\kappa_c=10^{-4}$, where
$\chi_l$ is the extinction coefficient of the line at the line center
and $\kappa_c$ is the local b-f absorption coefficient.  This typically
leads to about $10\alog{6}$ lines which are selected from master line
lists (with 47 million atomic and up to 350 million molecular lines).
The profiles of these lines are assumed to be depth-dependent Voigt or
Doppler profiles (for very weak lines). Details of the computation of the
damping constants and the line profiles are given in \cite{vb10pap}. We
have verified in test calculations that the details of the line
profiles and the threshold $\Gamma$ do not have a significant effect
on either the model structure or the synthetic spectra.  In addition,
we include abound 2000 photo-ionization cross sections for atoms and
ions \cite[]{mathisen84,verner95}.

\subsection{Molecular line data}

One of the most important improvements of the new models compared to
the \cite{MDpap} models (hereafter, AH95) is that new molecular line data  have recently
become available.  The first improvement of our molecular line list
is the addition of the HITRAN92 \cite[]{hitran92} database, first
incorporated in our ``Extended'' model grid (Allard \& Hauschildt,
unpublished).
For the molecules with lines available from
other sources (e.g., water vapor and CO) we prefer to use
other sources in model calculations because the \hitran\ lists
are fairly incomplete (they have been prepared for the conditions
of the Earth atmosphere) although the line data are of very high
quality. The OH molecule is an exception because the vibrational
bands present in the \hitran\ database are {\em not} present in the
Kurucz CD~15 \cite[]{cdrom15} dataset listing electronic transitions of OH.
Therefore, we use both \hitran\ and CD~15 OH line data simultaneously
in the model calculations.

We have replaced the vibration-rotation lines of CO \cite[]{cdrom15} used in
AH95 with the more recent CO line list calculated by Goorvich
\cite[]{GoorCOa,GoorCOb}. This list is more accurate than the old
CO line data. However, we have kept the electronic CO transitions
available on Kurucz's CD~15.
Similarly, we have replaced the Kurucz CN line list (CD~15)
with a more recent calculation by J{\o}rgensen \cite[]{CNlines}. 
CN lines are comparatively weak in most of the models presented here.
In addition, we have added line lists for YO \cite[]{yolines} and ZrO
\cite[]{zrolines} as well as H$_3^+$ partition functions and 
lines \cite[]{h3p,h3pl}.

Another improvement over the ``Base'' model grid \cite[]{MDpap} is the replacement
of the straight mean TiO opacities with the list of TiO lines 
computed by J{\o}rgensen \cite[]{TiOJorg}. This list includes a total
of more than 12 million lines for the most important band systems of
5 TiO isotopes. On  average, the opacity of TiO lines is about
a factor of 2 smaller than the old straight mean opacities. For any
given effective temperature this will result in weaker TiO features when
compared to the old straight means.

The absorption coefficient of water has been a long standing problem for
modelers of cool dwarf atmospheres. In our Base model grid we have used
the straight means of the \cite{ludwig} water opacity tables. These
are known to overestimate the water opacity significantly at higher
(and for M dwarfs more important) temperatures \cite[]{schryb94} but
they are accurate at lower temperatures. The currently available list
of water lines of Miller \& Tennyson \cite[\mtwater,][]{h2olet} is
incomplete and therefore some water opacity will be missing in models
with $\Teff<4000\K$.  Comparison with high-resolution observations in
spectral regions where the \mtwater\ list should be complete show that
its accuracy is very high \cite[]{mnwater}. In the models presented here,
we have used the \mtwater\ list.  Recently, a new water vapor line list
with more than 300 million lines was published by \cite{ames-water-new}.
A new and improved version of the Miller \& Tennyson list is currently
being calculated (Viti \etal, in preparation). In a subsequent publication
we will show the differences in the model structure and the synthetic
spectra for low effective temperature models calculated using different
water vapor line lists.

\subsection{The equation of state}

The equation of state (EOS) is an enlarged and enhanced version of the
EOS used in AH95. We include about 500 species (atoms, ions and molecules)
in the EOS. This set of EOS species was determined in test calculations.
The EOS calculations themselves follow the method discussed in AH95.
We did {\em not} include dust condensation in the calculations presented
here because it does not affect the models in the range of parameters
we consider. Models for effective temperatures lower than about
2500--$3000\K$ need to include these effects and will be discussed
in a subsequent paper.  Some of the molecular data were updated,
details are given in the aforementioned paper.

\section{Results}

\subsection{The Grid}

The NextGen model grid was calculated from $3000\K$
to $10000\K$ (in steps of 100--$200\K$) for $3.5 \le \logg \le 5.5$
(in steps of 0.5) and metallicities of $-4.0 \le \mh \le 0.0$ (in
steps of 0.5).  We use the recent solar abundances given in Table 5 of
\cite{jaschek95}. The changes in the abundances compared to the previous
\cite{solab89} data are in general small, with the exception of the
iron abundance which changed from $7.67$ to $7.53$. The grid contains a
total of 2142 models in this range of parameters (the complete NextGen
grid includes 3026 models).  In all models, convection is treated in
the mixing length approximation with the mixing length set to unity. We
will discuss the model grid for lower effective temperatures, the VLMS to
Brown Dwarf range of spectral class M, in a separate paper. For effective
temperatures higher than 7000--$10000\K$ NLTE effects become important,
so LTE models at higher effective temperature will become increasingly
unrealistic. In cooler models, NLTE effects are important for individual
atomic lines of, e.g., Ti~I \cite[]{tipap}, for effective temperatures
below about $4500\K$ but NLTE does not have significant effects on the
low resolution spectra or the structure of these models.  Therefore, we
include a few representative NLTE models (for $7000\le\Teff\le 10000\K$)
in the model grid presented here.  A full NLTE grid (with the complete
set of NLTE species available in \PHOENIX) is currently being constructed
and will be discussed in a later paper (Hauschildt \etal, in preparation).

\cite{epscma,betacma} have shown that for effective temperatures higher
than about $18000\K$ the combined effects of line blanketing and {\em
spherical geometry} are of crucial importance for main sequence stars with
$\logg \le 4.5$.  In this parameter range, plane parallel models do {\em
not} deliver enough EUV flux compared to spherical models or observed
EUV spectra.  However, for effective temperatures below about $10000\K$
the use of the plane parallel approximation in the model construction for
$\logg \ge 3.5$ does not result in significant differences to spherical
models. Therefore, we have used the plane parallel  geometry to calculate
the present grid.  A grid of NLTE spherical model atmospheres for OB
main sequence stars is currently being constructed and will be discussed
in a separate paper (Aufdenberg \etal, in preparation).  For effective
temperatures lower than about $3000\K$ the effects of dust formation
and/or dust opacity become important. This significantly changes the
physics of the model atmospheres and the formation of the spectrum,
therefore, we will discuss these models in a separate paper (Allard \etal,
in preparation).

The synthetic spectra and the model structures are available via
anonymous FTP from {\tt ftp://calvin.physast.uga.edu/pub/NextGen} or
via the WWW URL {\tt http://dilbert.physast.uga.edu/\tilt yeti} and
constitute about 560 MB of data. For each model, the model structure
(in the form of the \phoenix\ output file of the last model iteration)
and the synthetic low resolution spectra (directly the result
of the model iterations) are available. The model fluxes are given
as tables of $F_\lambda$ in erg/s/cm$^2$/cm versus wavelength in \AA\
to make comparisons with observed spectra and other model calculations
easier. The spectra are given on an semi-regular wavelength grid with
about 23,000 wavelength points from $1\ang$ to $1\,$cm, the actual
wavelength grid depends on the model.

\subsection{Comparison to Kurucz~92 models}

In Figs.~\ref{uv}--\ref{4000} we compare the \phoenix\ synthetic spectra with
\atlasIX\ spectra for a range of effective temperatures with fixed
solar abundances. Note that while Kurucz~92 uses the
\cite{solab89} solar abundances, we use the more recent solar abundances
given in Table 5 of \cite{jaschek95}. The changes in the abundances
will introduce small but systematic differences between the Kurucz~92
models and our  model structures and between the spectra because the
abundances of 
the key elements C, N, O, and Fe are different. The \phoenix\
spectra have been degraded by convolution
with a Gaussian kernel to more closely resemble the resolution of the
\atlasIX\ spectra. The spectra for the Kurucz models were obtained
using the {\tt IDL} routine {\tt KURGET1} and the corresponding database
of models available in the {\IUE} reduction and data analysis 
package {\tt IUERDAF}.

In general, our models agree well with the \atlasIX\ spectra for
effective temperatures higher than about $5000\K$.  We expect better
agreement between \phoenix\ and \atlasIX\ spectra for higher effective
temperatures since we have included the latest version of Kurucz atomic line
list into \phoenix\ and  atomic lines  dominate the overall opacity in
this temperature range. 
The remaining differences between the spectra are probably caused by
the slightly different abundances, different methods for treating the
line blanketing (ODF in \atlasIX\ versus direct opacity sampling in
\phoenix), the fact that the ODF's in the \atlasIX\ models were 
constructed using a previous version of the Kurucz line lists,
and by differences in the treatment of, e.g., scattering and
b-f and f-f opacities. These different treatments alter  the 
model structures, resulting in systematic variations between the spectra. 
In the ultraviolet (UV), the two model grid spectra differ in several
regions (cf.\ Fig.~\ref{uv}), which might  in part be due to the fact that
the \atlasIX\ spectra are based on ODF's. The general features of the
UV spectra are similar. Some of the features in the \phoenix\
spectrum are deeper than the \atlasIX\ features. This could  partly
be due to the different abundances and to changes in the atomic line list
itself since the calculation of the \atlasIX\ models.

We compare the models in the region of the H~I Balmer lines in
Fig.~\ref{balmer}. The lines widths agree well, however, the H~I lines
in the \phoenix\ spectrum seem to be a little deeper than the \atlasIX\
lines. This is probably simply an artifact of the ODF versus opacity
sampling treatment of the lines. In addition, we have inserted wavelength
points in the cores on the H~I lines. For comparison, we also show the
\phoenix\ spectrum at its original resolution with the fluxes scaled
by a factor of 0.5. The opacity sampling method results in many more
line features appearing in the spectrum, similar to \atlasXII\ results
shown in \cite{castelli94}.  Figure \ref{uv} compares our models to
Kurucz~92 for $\logg=4.0$, $\Teff=10000\K$ and for solar abundances. The
differences between the spectra are essentially the same over the whole
range of gravities and are therefore not shown. The \phoenix\ spectra
show less flux in some UV regions (e.g., around $1500\ang$) and the UV
lines are generally deeper than in the \atlasIX\ spectra. The flux that
is intercepted in the UV is redistributed into the Paschen continuum so
that the total energy flux through the atmosphere is given by $\sigma
\Teff^4$. Thus, the differences in the spectra are probably caused mainly
by a larger UV opacity in the opacity sampling \phoenix\ models.

For lower effective temperatures, the differences between the \phoenix\
and the \atlasIX\ spectra become larger. In Fig.~\ref{4000} we show a
more detailed comparison between models with $\Teff=4000\K$, $\logg=4.0$
and solar abundances. The figure clearly shows the spectral regions where
the two spectra do not agree well. In particular, the ``pseudo-continuum''
between about $0.9\,\mu$ and $1.6\mu$ lies at about a 10\% lower flux
level in the \atlasIX\ spectrum than in the \phoenix\ spectrum. In
the optical spectral region between $6000\ang$ and about $8000\ang$ the
\phoenix\ spectrum shows a number of molecular bands (due to TiO and VO?)
that are not seen in the \atlasIX\ spectrum. These bands are more important
in the VLMS and Brown Dwarf range of the NextGen grid. The Kurucz TiO
line list results in much weaker TiO features and different band shapes
than the J{\o}rgensen TiO list that we use \cite[]{andydipl}. This
is more important for lower effective temperatures and will cause more
significant deviations between the \phoenix\ and the \atlasIX\ models.

There are also discrepancies between the spectra in the near infrared
(NIR) spectral region. The CO bands are somewhat different, which is
probably caused by the ODF versus opacity sampling treatment of the line
blanketing of the calculations. The CO line data that we use
are very similar to the Kurucz CO line list \cite[]{andydipl}. We include
water lines in the model calculations and the spectral synthesis, which
is not included in the \atlasIX\ models. This causes the differences in
the molecular absorption band around $2.5\,\mu$, this feature is due to
a strong water vapor band. Therefore, it is likely that the differences for low
effective temperatures can be explained mainly by the different opacity
data used in \phoenix\ and \atlasIX. At effective temperatures as high
as $4000\K$, differences between the number of molecules and the
molecular data used in the EOS's of \phoenix\ and \atlasIX\
are probably unimportant. At lower temperatures, however, the larger 
EOS of \phoenix\ will likely increase the differences
between the two sets of models.

\subsection{Comparison to NLTE models}

 In order to investigate the importance of NLTE effects on the
models presented in this paper, we have calculated small number of
representative NLTE models with parameters roughly appropriate for the
Sun and Vega ($\alpha$ Lyr). We include a large number of levels and
lines in NLTE in order to make the model calculations as realistic as
possible. In addition, species that we do not include explicitly
in NLTE are included in the LTE approximation in order to 
describe their effects on the model structure and the spectrum,
see \cite{tipap} and \cite{epscma} and references therein for 
details of the method. This allows us to determine which models 
require NLTE and for which models LTE provides an
adequate description of line formation. In the outer, optically thin,
regions of the atmosphere, NLTE effects will always become important
(even in the absence of a chromospheric temperature rise), but for models
of stellar photospheres, it is crucial to know if NLTE effects are
important in the line forming regions. We have computed detailed NLTE
models for two cases:  for relatively cool, G-type stars we chose model
parameters approximately appropriate for the Sun; and for hotter stars,
where we chose model parameters appropriate for Vega. The models serve 
to demonstrate the effects of NLTE on the structure and spectra, we did
not make any attempt to produce detailed fits to either the Sun or Vega.
The models include a large number of NLTE species and use 
the same version of \phoenix\ and the same input physics
as the NextGen models.

\subsubsection{Solar-Type Stars}

For the solar-type model, we used $\Teff=5770\K$, $\logg=4.44$ and
solar abundances. The only difference with the NextGen models is the use
of \phoenix's NLTE mode, which includes the effects of multi-level 
NLTE on the EOS, the radiative transfer (both lines and continua)
and the structure of the atmosphere. This model is a very simple 
model for the solar spectrum, for example we have not used the best
value for the mixing length parameter for the sun of $\alpha=1.5$ and
have used only simple approximations for the damping constants of
most spectral lines. Therefore, the model cannot reproduce the solar
spectrum as well as a fine tuned model could, we only use it to 
demonstrate the effects of NLTE on solar type stars.

Table~\ref{nltetab} gives a complete listing of {\em all} NLTE species
currently available in \phoenix. For the solar model, not all of these are
important, therefore we have used only the following sub-set of species in
NLTE: H~I, He~I--II, Li~I, Na~I, Ne~I, Mg~II, Ca~II, Ti~I--II, S~II--III,
Si~II--III, C~I--II, N~I--II, O~I--II, and Fe~I--III.  This resulted in
a total of 4,143 NLTE levels and 49,324 primary NLTE transitions. 26,786
of the primary NLTE lines were treated with detailed Voigt profiles,
the remaining 22,538 weak primary NLTE lines were treated with Gaussian
profiles to save CPU time. In test calculations we verified that this
does not change the results significantly. In addition, the NLTE models
include 218,009 secondary NLTE lines as well 385,484 LTE background atomic
lines and 1,566,441 LTE molecular lines, these lines were selected using
the same criterion that we use for the LTE models. The calculation was
performed with a variable resolution wavelength grid with about 300,000
points, including extra points that are inserted to resolve NLTE lines.
We are currently adding Mg I, Si I, S I, Ca I, and H$^-$ to our list of
NLTE species and we will produce improved models in cases where these
species are significant.

In Fig.~\ref{solar-global} we show the comparison between the Kitt Peak
Solar Atlas \cite[]{kpsun} spectrum and the simplified solar model. The resolution of
the Kitt Peak spectrum is extremely high ($0.01\ang$). Therefore, we have
reduced the resolution by convolving the spectrum with a
Gaussian kernel with a width of $10\ang$, shown
as the full curve in Fig.~\ref{solar-global}. The same procedure was used
to plot the synthetic spectrum (dotted line). In Fig.~\ref{solar-details}
we show the comparison at an enlarged wavelength scale. In this figure,
the Kitt Peak spectrum and the \phoenix\ spectrum were left at their 
original resolution (the resolution of the \phoenix\ spectrum is
variable, $\Delta \lambda \le 0.05\ang$). The fit is in general good,
although fine tuned model parameters would improve the comparison. 

The structure of the atmosphere is shown in Fig.~\ref{sun-struc} and in
more detail in Fig.~\ref{sun-struc2}. The differences between the LTE and
NLTE structures is very small.  We show selected departure coefficients
for the solar model in Fig.~\ref{sun-bi}. In the line forming regions,
typically between $\tstd=10^{-3}$ and unity, the departure coefficients
are very close to one and many lines thus form basically under LTE
conditions. The departure coefficients drop rapidly at lower optical
depths. In this region, the chromospheric temperature rise (which is
{\em not} included in the model) would change the results drastically
above the solar temperature minimum of about $4400\K$. The wide range
of different atoms and ions that are included in this model gives an
overall impression of the average importance of the NLTE effects in the
photosphere of the Sun. We expect that the other elements, not treated
in NLTE in this model, will roughly follow the selected ions shown in
Fig.~\ref{sun-bi}. The NLTE effects on the photospheric structure and
the line forming regions are small, therefore, LTE models can be used
as a
first approximation to model the photospheres of solar type stars. This
is consistent with the results of \cite{anderson89}, who used a different
numerical approach.   The relatively high electron
densities in the line forming region, which make the collisional rates
more important in solar type stars compared to either hotter or cooler stars
is the reason that LTE is a good approximation for solar type stars.

\subsubsection{Vega ($\alpha$ Lyr)}

The detailed analysis of the spectrum of Vega with \atlasIX\ and
\atlasXII\ models has been discussed by \cite{castelli94}. Here, we
compare \phoenix\ LTE and NLTE models computed with the parameters
found for Vega by \cite{castelli94} to the observed Vega spectrum
to show the effects of NLTE on the spectral energy distribution for
stars with Vega's effective temperature.  We have computed an
NLTE model with 
$\Teff=9550\K$, $\logg=3.95$ and the ``Vega abundances'' given in table
4 of \cite{castelli94}. We have used
the following species in NLTE: H~I, He~I--II, Ne~I, Mg~II,
Ca~II, Ti~I--II, S~II--III, Si~II--III, C~I--II, N~I--II, O~I--II,
and Fe~I--III, which includes the most important species for the
conditions in Vega.  This resulted in a total of 3,787 NLTE levels and 47,593
primary NLTE transitions. 25,449 of the primary NLTE lines were treated
with detailed Voigt profiles, the remaining 22,144 weak primary NLTE
lines were treated with Gaussian profiles to save CPU time. In addition,
the NLTE models include 223,563 secondary NLTE lines as well 219,774 LTE
background lines, these lines were selected using the same
criterion that we use for the LTE models. The calculation was performed
with a variable resolution  wavelength grid with 204,434 points, including
extra points that are inserted to resolve NLTE lines.

We show a comparison between the NLTE model and Vega observational data in
Fig.~\ref{vega-special}. In general, the fit is good, no attempts
have been made to fine-tune some of the abundances (e.g., to reduce the 
too strong absorption feature just shortward of $1600\ang$). A more detailed
analysis of the Vega data will be presented elsewhere (Aufdenberg \etal, in
preparation).

We compare the NLTE and LTE structures for a Vega model with
$\Teff=9500\k$ in Fig.~\ref{vega-struc} and the departure
coefficients of the ground states of some selected ions are shown in
Fig.~\ref{vega-bi}. The departures from LTE are larger for the Vega model
than for the solar model shown above. The changes in the temperature
structure are not very large in the deeper layers, but can reach nearly
$1000\K$ in the outermost, optically thin, zones of the model.
In the line forming regions, NLTE effects are generally small so
that the spectra of the NLTE and the LTE models are not very different.
Since the departures from LTE are larger for  Vega than for the solar
model,
the validity of the assumption of LTE is weaker for Vega than for the
solar model.
This is the main reason why the LTE models that we present in
this paper essentially stop at $\Teff=10000\k$, for higher effective
temperatures NLTE effects will rapidly become important.

\section{Summary and Conclusions}

In this paper we presented a new set of models for main sequence 
stars for effective temperatures from $3000\K$ to $10000\K$. We include
mostly LTE models, however, NLTE effects become progressively important
for effective temperatures larger than about $7000\K$. Therefore, we have
included self-consistent NLTE models calculated with a subset
of the NLTE species that are available in \phoenix. 

We find that our models agree well with the Kurucz~92 grid for solar
type stars ($5000\le \Teff\le 7000\K$) and that our LTE models for
stars with effective temperatures up to about $10000\K$ also agree
reasonably well with the Kurucz models. For lower temperatures we
find some differences between the model grids. The reason for this is
that our calculations are designed to include the atmospheres of very
low mass stars and brown dwarfs, so that they include a more detailed
molecular EOS and a better set of molecular opacity sources than the
Kurucz~92 models. For higher effective temperatures, NLTE effects become
important. Our results for Vega show that NLTE has an effect on the
structure of these models. Therefore, detailed NLTE models are required
for effective temperatures higher than about $7000\K$. Furthermore,
\cite{epscma,betacma} have shown that the combined effects of line
blanketing and spherical geometry are essential ingredients in model
atmosphere models for effective temperatures  $\ga 18000\K$.  We will
present spherical NLTE models for larger effective temperatures in a
subsequent paper (Aufdenberg \etal, in preparation).

 It appears that the assumption of LTE is a useful approximation
only in a very narrow range of effective temperatures near that
of solar type stars. For cooler and even more so, for hotter stars,
NLTE effects are important and should be included in model atmosphere
calculations.

\acknowledgments
We thank Jason Aufdenberg, David Alexander, Robert Kurucz, and Sumner
Starrfield for helpful discussions. This work was supported in part by NSF
grant AST-9720704, NASA ATP grant NAG 5-3018 and LTSA grant NAG 5-3619 to the
University of Georgia,  by NSF grant AST-9417242,  NASA grant NAG5-3505 and an
IBM SUR grant to the University of Oklahoma, and NASA LTSA grant NAG5-3435 to
Wichita State University.  The NSO/Kitt Peak FTS data of the solar spectrum
used in this paper were produced by NSF/NOAO.  Some of the calculations
presented in this paper were performed on the IBM SP2 and the SGI Origin 2000
of the UGA UCNS, at the San Diego Supercomputer Center (SDSC) and the Cornell
Theory Center (CTC), with support from the National Science Foundation, and at
the NERSC with support from the DoE. We thank all these institutions for a
generous allocation of computer time.

\clearpage

\bibliography{yeti,opacity,mdwarf,radtran,general,opacity-fa}

\clearpage

\section{Tables}

\begin{table}
\caption[]{\label{nltetab}Complete listing of {\tt PHOENIX} (version 8)
NLTE species.  The tables entries are of the form N/L, where N is the
number of NLTE levels and L the number of primary NLTE lines for each
model atom. The calculations presented in this paper include all species 
listed in the table with the exceptions of Li, Ti, and Co. PETER: you
say you use Li and Ti in the sun and Ti in Vega.}
\typeout{Fix table caption}
\smallskip
\begin{tabular}{*{7}{l}}
\hline
\hline
        &   I  & II & III & IV & V & VI \\
\hline
H       &  30/435\\
He      &  11/14 & 15/105\\
Li      &  57/333 \\
C       &  228/1387 & 85/336 & 79/365 & 35/171 \\
N       &  252/2313 & 152/1110 & 87/266 & 80/388 & 39/206 &15/23 \\
O       &  36/66 & 171/1304 & 137/766 & 134/415 & 97/452 & 39/196 \\
Ne      &  26/37 \\
Na      &  3/2 \\
Mg      &  &  18/37 \\
Si      &  & 93/436 & 155/1027 \\
S       &  & 84/444 & 41/170 \\
Ca      &  & 87/455 \\
Ti      & 395/5279 & 204/2399 \\
Fe      & 494/6903 & 617/13675 & 566/9721\\
Co      & 316/4428 & 255/2725 & 213/2248\\
\hline
total:  & & 5346/60637 \\
\hline
\hline
\end{tabular}
\end{table}

\clearpage
\section{Figures}


\begin{figure}[t]
\caption[]{\label{uv} Comparison between \phoenix\ (full line) and
\atlasIX\ (dotted line) model spectra in the UV for $\Teff=10000\K$,
$\logg=4.0$ and solar abundances. The \phoenix\ spectrum has been smoothed
with a Gaussian kernel to $10\ang$ resolution.}
\end{figure}

\begin{figure}[t]
\caption[]{\label{balmer} Comparison between \phoenix\ (full line) and
\atlasIX\ (dotted line) model spectra in the UV for $\Teff=10000\K$,
$\logg=4.0$ and solar abundances. The \phoenix\ spectrum has been
smoothed with a Gaussian kernel to $20\ang$ resolution. We also show the
\phoenix\ spectrum at its original resolution scaled by a factor of 0.5
for comparison.}
\end{figure}


\begin{figure}[t]
\caption[]{\label{4000} Comparison between \phoenix\ (full line) and
\atlasIX\ (dotted line) model spectra for $\Teff=4000\K$, $\logg=4.0$
and solar abundances. The \phoenix\ spectrum was convolved with a Gaussian
kernel of $20\ang$ width for wavelengths smaller than $1\,\mu$m and with a
Gaussian kernel of $50\ang$ width for wavelengths larger than $1\,\mu$m.}
\end{figure}

\begin{figure}[t]
\caption[]{\label{sun-struc} Comparison between the NLTE and LTE
structure of \phoenix\ solar 
NLTE models. $\tstd$ is the optical depth in the continuum at $1.2\mu$.}
\end{figure}

\begin{figure}[t]
\caption[]{\label{solar-global} Comparison between the Kitt Peak Solar Atlas
spectrum (thick curve) and the solar NLTE model (thin curve),
both reduced to a resolution of $\Delta\lambda=10\ang$. No attempts were made
to fine-tune the model. Terrestrial features (e.g.,
at about 0.75, 0.95 and $1.12\,\mu$m) were not removed from the 
observed solar spectrum.}
\end{figure}

\begin{figure}[t]
\caption[]{\label{solar-details} Comparison between the Kitt Peak Solar
Atlas spectrum (thick curves) and the 
solar NLTE model (thin curves) for selected
wavelength regions. Both spectra were smoothed to a resolution
of $\Delta\lambda=0.05\ang$. No attempts were made to fine-tune the model.}
\end{figure}

\begin{figure}[t]
\caption[]{\label{sun-struc2} Comparison between the NLTE and LTE
structure of \phoenix\ solar 
NLTE models as function of the height above $\tstd=1$}.
\end{figure}

\begin{figure}[t]
\caption[]{\label{sun-bi} Departure coefficients of the ground states
of selected ions as function of 
$\tstd$ (the optical depth in the continuum at $1.2\mu$) for the solar model
with $\Teff=5770\K$ and $\logg=4.44$.}
\end{figure}

\begin{figure}[t]
\caption[]{\label{vega-special} Comparison between a \phoenix\ Vega 
NLTE model (dotted/thick line)
and Vega absolute spectrophotometry and IUE observational data. The model
parameters are given in the figure, the ``Vega abundances'' are taken
from \cite{castelli94}. The model spectrum has been convolved with a
Gaussian kernel of $6\ang$ half-width to facilitate comparison with the 
low resolution data.}
\end{figure}

\begin{figure}[t]
\caption[]{\label{vega-bi} Departure coefficients of the ground states
of selected ions as function of 
$\tstd$ (the optical depth in the continuum at $1.2\mu$) for a Vega model
with $\Teff=9500\K$ and $\logg=3.95$.}
\end{figure}

\begin{figure}[t]
\caption[]{\label{vega-struc} Comparison between the NLTE and LTE
structure of \phoenix\ Vega 
NLTE models. $\tstd$ is the optical depth in the continuum at $1.2\mu$.}
\end{figure}

\end{document}